\documentclass[twocolumn,preprintnumbers,amsmath,amssymb]{revtex4-1}
 \pdfoutput=1
\usepackage{color}% color text
\usepackage{graphicx}
\usepackage{amssymb}
\usepackage{amsmath}
\usepackage{mathrsfs}
\usepackage{subfigure}

\begin{document}

\title{Triply-resonant Continuous Wave Parametric Source with a Microwatt Pump}

\author{A. Martin$^{1,2,\dagger}$, G. Moille$^{2,\dagger}$,  S. Combri\'e$^2$,  G. Lehoucq$^2$, T. Debuisschert$^{2}$, J. Lian$^3$, S. Sokolov$^3$, A. P. Mosk$^3$ and A. de Rossi$^{2}$}
\email{alfredo.derossi@thalesgroup.com}
\address{
$^1$Laboratoire de Photonique et de Nanostructures, CNRS, Universit\'e Paris-Saclay, 91460 Marcoussis, France \\
$^2$Thales Research and Technology, 91767 Palaiseau, France\\
$^3$Light in Complex Systems (COPS), Debye Institute for Nanomaterials Science, University of Utrecht, 3584 CC Utrecht, The Netherlands\\
$^\dagger$equal contribution}

% 	-- Abstract --
%----------------------------------------------------------------------------------------------
\begin{abstract}
We demonstrate a nanophotonic parametric light source with a record high normalized conversion efficiency of $3\times 10^6\, W^{-2}$, owing to resonantly enhanced four wave mixing in coupled high-Q photonic crystal resonators. The rate of spontaneously emitted photons reaches 14 MHz. 
\end{abstract}
%--	End Abstract -----------------------------------------------------------------------------

\maketitle
% 	-- Section 1 : Introduction --
%----------------------------------------------------------------------------------------------
\section{Introduction}
Nonlinear frequency conversion processes are attractive for the generation of light at arbitrary wavelengths. Integrated photonic devices based on four-wave mixing (FWM) are of particular interest due to the promise of compact all-optical sources for signal processing metrology and quantum information transmission. 
In a resonator with a non-depleted pump the efficiency of frequency conversion due to four wave mixing (FWM) scales as $Q^4$\cite{azzini2012}, provided the pump, the probe and the idler are all spectrally matched with a resonance. The generation of photon pairs by spontaneous frequency conversion is similarly enhanced \cite{Grassani2015,azzini2013,matsuda2012}.\\
The efficiency of frequency conversion also scales as $1/V_{mode}^4$ (where $V_{mode}$ is the modal volume of the resonator), therefore small resonators are favored. Particularly, photonic crystal (PhC) cavities, with a modal volume of order $\lambda^3$, are expected to provide ultimate conversion efficiency.
It is however notoriously difficult to obtain equally spaced resonances by coupling photonic crystal resonators. Recently, Matsuda and co-workers have demonstrated efficient FWM and, consequently, photon pair generation, in a photonic crystal structure entailing 200 coupled resonators \cite{matsuda2012}. Azzini \textit{et al.}\cite{azzini2013} have demonstrated resonantly enhanced FWM and photon pair generation in a device consisting of 3 coupled PhC cavities. Here we report parametric amplification with 
high conversion efficiency at ultra low power which is favourable for scaling up. The rate of spontaneously emitted photons is also high. 
% -------------------- figure 1 -----------------------------
    \begin{figure}[h]
      \centering
              \includegraphics[width=0.9\columnwidth]{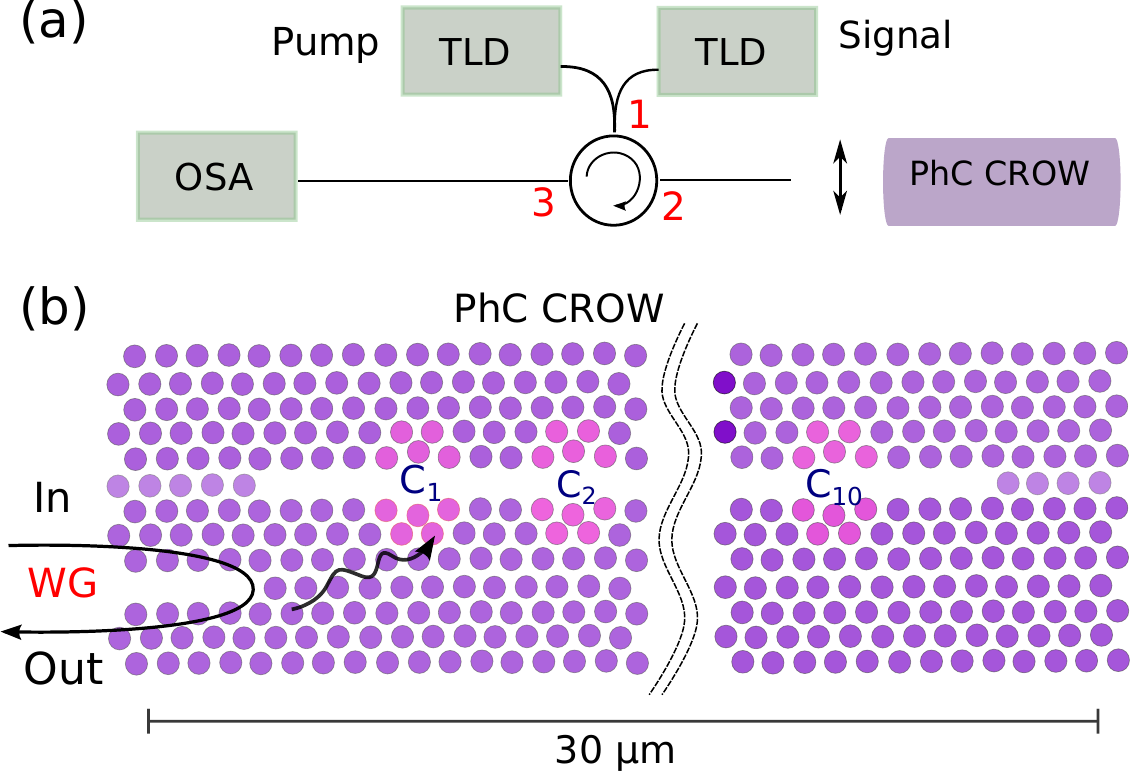}
       \caption{(a) Measurement setup: TLS tunable laser source, OSA Optical Spectrum Analyser;(b) device layout.}
              \label{fig:fig_1}
    \end{figure}
% -------------------- figure 1 -----------------------------
%
% 	-- Section 2 : Results and discussion --
%----------------------------------------------------------------------------------------------
\section{Results and discussion}
\begin{figure}
\centering
\includegraphics[width=0.9\columnwidth]{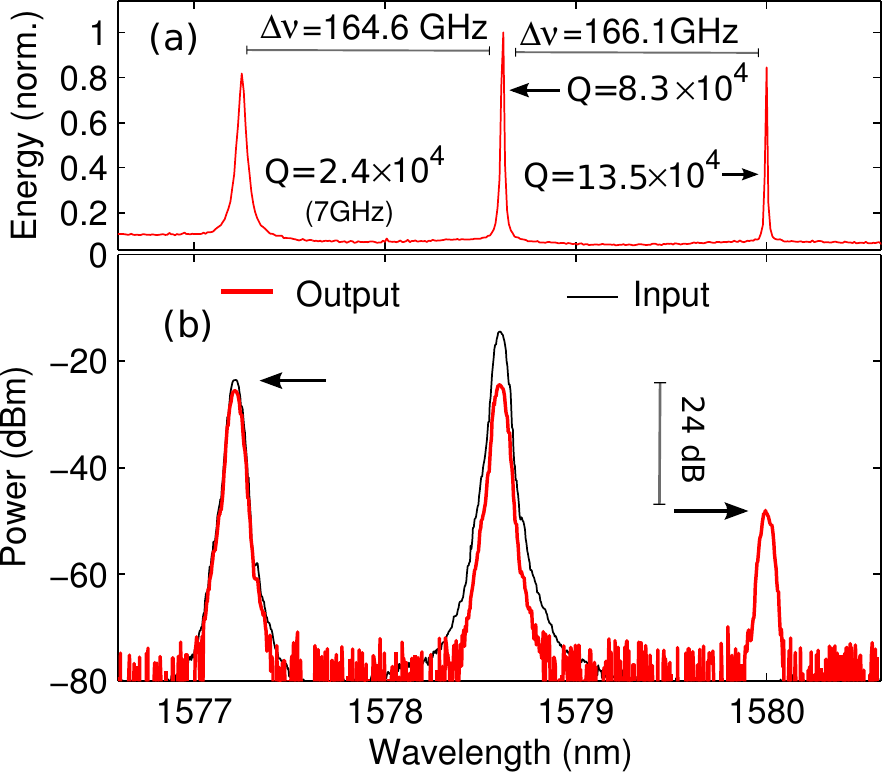}
\caption{(a) linear scattering spectrum ; (b) input and output (reflected) spectra.}
\label{fig:fig_2}
\end{figure}

Our nonlinear resonator is made of a 180 nm thin suspended membrane of Gallium Indium Phosphide (GaInP). It is represented schematically in Figure \ref{fig:fig_1}b below and it consists of a chain of 10 coupled cavities obtained by modulating the width of a photonic crystal waveguide, similar to ref.\cite{matsuda2012}. The linear scattering spectrum is obtained by measuring the optical signal leaking out of the resonator with an infrared camera. This is shown in Fig. \ref{fig:fig_2}a. Three peaks (out of 10) have been identified such that their frequency spacing is almost identical: 166.1 and 164.6 GHz. Importantly, the residual difference is smaller than the spectral width (7GHz) of the broadest resonance, which has $Q$$=$$2.4\times 10^4$. This equal spacing enables phase-matched resonant-enhanced FWM. The other two resonances have $Q$$=$$8.3\times 10^4$ and $13.5\times 10^4$, leading to an averaged Q factor of  $(Q_s Q_i Q_p^2)^{1/4}=7\times 10^4$.

Light is injected in the system through a side photonic crystal waveguide with slightly larger width, from which it  couples evanescently to the resonators. The pump and the idler are generated with two tunable lasers which are  combined on a beamsplitter and coupled into the waveguide using an objective lens. A fiber circulator is used to extract the  signal propagating backwards from the sample. The total insertion losses (circulator, the objective lens , the side PhC waveguide) amount to 11 dB. Because of the symmetry of the system the on-chip power levels 
can be estimated accurately by applying 5.5 dB offset to the measurements before the port 1 and after the port 3 of the circulator.

\begin{figure}[h]
\centering
\includegraphics[width=\columnwidth]{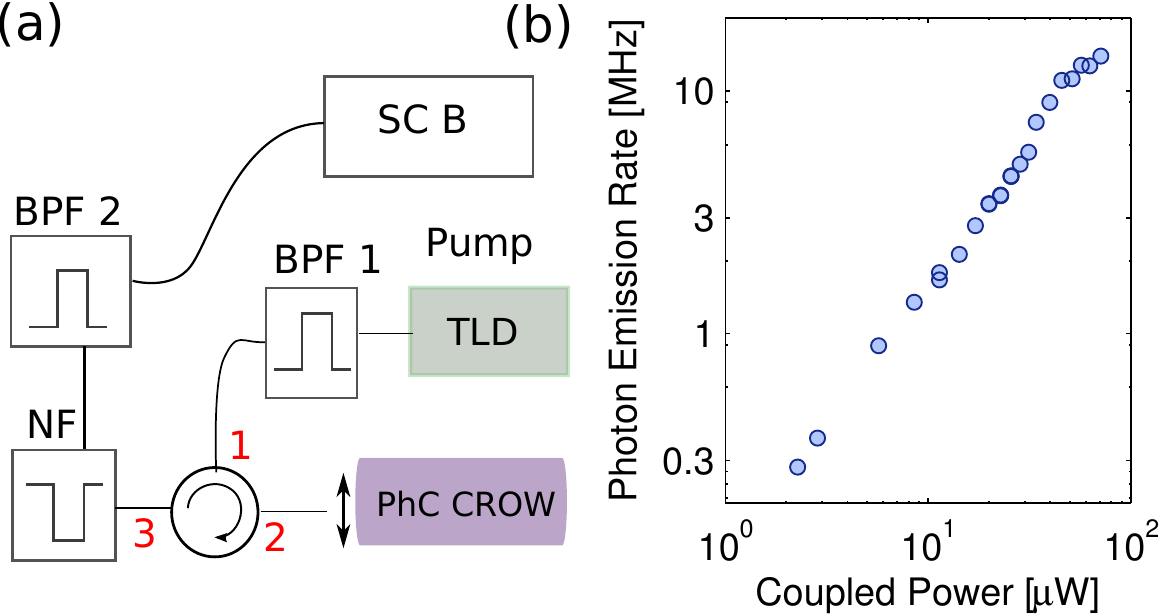}
\caption{(a) Photon count setup: BPF band pass filterm, NF Notch Filter, SC B Superconductive Bolometer ; (b) Rate of spontaneously emitted photons vs. coupled pump  power.}
\label{fig:fig_3}
\end{figure}

The FWM signal is detected in CW using an Optical Spectrum Analyzer (OSA). The idler signal is easily detected by the OSA as the pump and the signal are properly tuned. The corresponding input and the output spectra are shown in \ref{fig:fig_1}a. The power levels measured by the OSA have been calibrated with a power meter and then rescaled to on-chip levels.\\
The conversion efficiency is defined as the ratio of the power generated in the idler to the signal power level before the resonator: $\eta=P_{idler}/P_{signal,in}$. We detect an idler power of $15.7 nW$ from $4.2 \mu W$ signal input, evaluated \textit{on-chip}. This yields a conversion efficiency of -24 dB at a pump power of 36 $\mu W$.
The conversion efficiency is compared with recent experiments in ring resonators and coupled photonic crystal cavities (table \ref{table:1}). When necessary, it is estimated by converting the reported photon count rate into idler power. We observe a conversion efficiency that is  about 3 orders of magnitude larger than in ref. \cite{azzini2013}. This higher conversion efficiency follows from the fact that the averaged Q-factor of our cavities is an order of magnitude higher than in ref. \cite{azzini2013}.
Even compared to ring resonators with a Q-factor similar to our cavities, we find that the efficiency is large,  especially when normalized by the input power.
%	-------------------- table -----------------------------
\begin{table}
 \centering \caption{FWM in nanophotonic devices}
	\begin{tabular}{ccccccc}
    \hline
    Techn. & Material & Q-factor & $P$  & $\eta$  & $\eta_{norm}$ & Ref.\\
       &  &  & \textit{$\mu$W} & \textit{dB} &  $W^{-2}$ &\\
    \hline
    Ring res. & Si & $ 6\times10^4$ & 700  & -16.4 &$4.5\times 10^4$ &\cite{strain2015}\\
    3 PhC cav. & Si & $4\times10^3$  & 60 & -55 & $0.9\times 10^3$ & \cite{azzini2013}\\
    PhC CROW & GaInP & $7\times10^4$ & 36 & -24 & $3\times 10^6$ \\        
    \hline
   \end{tabular}
   \label{table:1}
\end{table}
%	-------------------- table -----------------------------
\\In addition we operated the device in the spontaneous emission regime. As shown in Fig.~\ref{fig:fig_3}a, two narrowband filters (BP1,BP2, Yenista, 60 dB rejection, adjusted to 100 pm width) are used to remove residual noise from the source and to filter the emitted photons. An additional notch filter (NF, Yenista, 50 dB rejection) is used at output to filter the pump out. A superconductive bolometer, Scontel, calibrated with an attenuated laser, is used for photon counting. The photon generation rate is extracted from the raw count rate after considering the insertion losses (coupling and filters) and the detection efficiency of the detector (fig. \ref{fig:fig_3}b). Measurement have been performed with NBF2 tuned either on the resonance in the blue side (1577.5nm) or the red side (1580 nm). A maximum ratio of 14 MHz is achieved when NBF 2 is tuned on the blue side, hence we estimate the normalized count rate to be larger than 200 MHz/nm.
This fairly large value of the emission rate is consistent with the fact that the ratio between spontaneous emission (photon pair generation) and stimulated emission (parametric frequency conversion) only depends on the Q-factor and the input signal power\cite{azzini2012}.
\section*{Conclusions}
In conclusion, we report very efficient four wave mixing  in a triply resonant photonic crystal structure. Owing to the large Q factor of the three resonances, the conversion efficiency is as high as -24 dB with only 36 $\mu W $of optical pump. The emission rate of spontaneous photons is 14 MHz, corresponding to a very high spectral density of $2 \times 10^8$ /s/nm as a result of the very narrow width of the cavity resonance.
\section*{acknowledgements}
This work was supported by the European Research Council (ERC), project PHAROS (grant 279248, P.I A. P. Mosk) and the FET-Open GOSPEL Project (grant 219299). A. Martin aknowledges support from IDEX Paris Saclay - IDI 2013.
%
% 	-- Bibliography --
%----------------------------------------------------------------------------------------------
\bibliography{biblio}

\begin{thebibliography}{5}%
\makeatletter
\providecommand \@ifxundefined [1]{%
 \@ifx{#1\undefined}
}%
\providecommand \@ifnum [1]{%
 \ifnum #1\expandafter \@firstoftwo
 \else \expandafter \@secondoftwo
 \fi
}%
\providecommand \@ifx [1]{%
 \ifx #1\expandafter \@firstoftwo
 \else \expandafter \@secondoftwo
 \fi
}%
\providecommand \natexlab [1]{#1}%
\providecommand \enquote  [1]{``#1''}%
\providecommand \bibnamefont  [1]{#1}%
\providecommand \bibfnamefont [1]{#1}%
\providecommand \citenamefont [1]{#1}%
\providecommand \href@noop [0]{\@secondoftwo}%
\providecommand \href [0]{\begingroup \@sanitize@url \@href}%
\providecommand \@href[1]{\@@startlink{#1}\@@href}%
\providecommand \@@href[1]{\endgroup#1\@@endlink}%
\providecommand \@sanitize@url [0]{\catcode `\\12\catcode `\$12\catcode
  `\&12\catcode `\#12\catcode `\^12\catcode `\_12\catcode `\%12\relax}%
\providecommand \@@startlink[1]{}%
\providecommand \@@endlink[0]{}%
\providecommand \url  [0]{\begingroup\@sanitize@url \@url }%
\providecommand \@url [1]{\endgroup\@href {#1}{\urlprefix }}%
\providecommand \urlprefix  [0]{URL }%
\providecommand \Eprint [0]{\href }%
\providecommand \doibase [0]{http://dx.doi.org/}%
\providecommand \selectlanguage [0]{\@gobble}%
\providecommand \bibinfo  [0]{\@secondoftwo}%
\providecommand \bibfield  [0]{\@secondoftwo}%
\providecommand \translation [1]{[#1]}%
\providecommand \BibitemOpen [0]{}%
\providecommand \bibitemStop [0]{}%
\providecommand \bibitemNoStop [0]{.\EOS\space}%
\providecommand \EOS [0]{\spacefactor3000\relax}%
\providecommand \BibitemShut  [1]{\csname bibitem#1\endcsname}%
\let\auto@bib@innerbib\@empty
%</preamble>
\bibitem [{\citenamefont {Azzini}\ \emph {et~al.}(2012)\citenamefont {Azzini},
  \citenamefont {Grassani}, \citenamefont {Galli}, \citenamefont {Andreani},
  \citenamefont {Sorel}, \citenamefont {Strain}, \citenamefont {Helt},
  \citenamefont {Sipe}, \citenamefont {Liscidini},\ and\ \citenamefont
  {Bajoni}}]{azzini2012}%
  \BibitemOpen
  \bibfield  {author} {\bibinfo {author} {\bibfnamefont {S.}~\bibnamefont
  {Azzini}}, \bibinfo {author} {\bibfnamefont {D.}~\bibnamefont {Grassani}},
  \bibinfo {author} {\bibfnamefont {M.}~\bibnamefont {Galli}}, \bibinfo
  {author} {\bibfnamefont {L.~C.}\ \bibnamefont {Andreani}}, \bibinfo {author}
  {\bibfnamefont {M.}~\bibnamefont {Sorel}}, \bibinfo {author} {\bibfnamefont
  {M.~J.}\ \bibnamefont {Strain}}, \bibinfo {author} {\bibfnamefont
  {L.}~\bibnamefont {Helt}}, \bibinfo {author} {\bibfnamefont {J.}~\bibnamefont
  {Sipe}}, \bibinfo {author} {\bibfnamefont {M.}~\bibnamefont {Liscidini}}, \
  and\ \bibinfo {author} {\bibfnamefont {D.}~\bibnamefont {Bajoni}},\
  }\href@noop {} {\bibfield  {journal} {\bibinfo  {journal} {Optics letters}\
  }\textbf {\bibinfo {volume} {37}},\ \bibinfo {pages} {3807} (\bibinfo {year}
  {2012})}\BibitemShut {NoStop}%
\bibitem [{\citenamefont {Grassani}\ \emph {et~al.}(2015)\citenamefont
  {Grassani}, \citenamefont {Azzini}, \citenamefont {Liscidini}, \citenamefont
  {Galli}, \citenamefont {Strain}, \citenamefont {Sorel}, \citenamefont
  {Sipe},\ and\ \citenamefont {Bajoni}}]{Grassani2015}%
  \BibitemOpen
  \bibfield  {author} {\bibinfo {author} {\bibfnamefont {D.}~\bibnamefont
  {Grassani}}, \bibinfo {author} {\bibfnamefont {S.}~\bibnamefont {Azzini}},
  \bibinfo {author} {\bibfnamefont {M.}~\bibnamefont {Liscidini}}, \bibinfo
  {author} {\bibfnamefont {M.}~\bibnamefont {Galli}}, \bibinfo {author}
  {\bibfnamefont {M.~J.}\ \bibnamefont {Strain}}, \bibinfo {author}
  {\bibfnamefont {M.}~\bibnamefont {Sorel}}, \bibinfo {author} {\bibfnamefont
  {J.~E.}\ \bibnamefont {Sipe}}, \ and\ \bibinfo {author} {\bibfnamefont
  {D.}~\bibnamefont {Bajoni}},\ }\href@noop {} {\bibfield  {journal} {\bibinfo
  {journal} {Optica}\ }\textbf {\bibinfo {volume} {2}},\ \bibinfo {pages} {88}
  (\bibinfo {year} {2015})}\BibitemShut {NoStop}%
\bibitem [{\citenamefont {Azzini}\ \emph {et~al.}(2013)\citenamefont {Azzini},
  \citenamefont {Grassani}, \citenamefont {Galli}, \citenamefont {Gerace},
  \citenamefont {Patrini}, \citenamefont {Liscidini}, \citenamefont {Velha},\
  and\ \citenamefont {Bajoni}}]{azzini2013}%
  \BibitemOpen
  \bibfield  {author} {\bibinfo {author} {\bibfnamefont {S.}~\bibnamefont
  {Azzini}}, \bibinfo {author} {\bibfnamefont {D.}~\bibnamefont {Grassani}},
  \bibinfo {author} {\bibfnamefont {M.}~\bibnamefont {Galli}}, \bibinfo
  {author} {\bibfnamefont {D.}~\bibnamefont {Gerace}}, \bibinfo {author}
  {\bibfnamefont {M.}~\bibnamefont {Patrini}}, \bibinfo {author} {\bibfnamefont
  {M.}~\bibnamefont {Liscidini}}, \bibinfo {author} {\bibfnamefont
  {P.}~\bibnamefont {Velha}}, \ and\ \bibinfo {author} {\bibfnamefont
  {D.}~\bibnamefont {Bajoni}},\ }\href@noop {} {\bibfield  {journal} {\bibinfo
  {journal} {Applied Physics Letters}\ }\textbf {\bibinfo {volume} {103}},\
  \bibinfo {pages} {031117} (\bibinfo {year} {2013})}\BibitemShut {NoStop}%
\bibitem [{\citenamefont {Matsuda}\ \emph {et~al.}(2012)\citenamefont
  {Matsuda}, \citenamefont {Le~Jeannic}, \citenamefont {Fukuda}, \citenamefont
  {Tsuchizawa}, \citenamefont {Munro}, \citenamefont {Shimizu}, \citenamefont
  {Yamada}, \citenamefont {Tokura},\ and\ \citenamefont
  {Takesue}}]{matsuda2012}%
  \BibitemOpen
  \bibfield  {author} {\bibinfo {author} {\bibfnamefont {N.}~\bibnamefont
  {Matsuda}}, \bibinfo {author} {\bibfnamefont {H.}~\bibnamefont {Le~Jeannic}},
  \bibinfo {author} {\bibfnamefont {H.}~\bibnamefont {Fukuda}}, \bibinfo
  {author} {\bibfnamefont {T.}~\bibnamefont {Tsuchizawa}}, \bibinfo {author}
  {\bibfnamefont {W.~J.}\ \bibnamefont {Munro}}, \bibinfo {author}
  {\bibfnamefont {K.}~\bibnamefont {Shimizu}}, \bibinfo {author} {\bibfnamefont
  {K.}~\bibnamefont {Yamada}}, \bibinfo {author} {\bibfnamefont
  {Y.}~\bibnamefont {Tokura}}, \ and\ \bibinfo {author} {\bibfnamefont
  {H.}~\bibnamefont {Takesue}},\ }\href@noop {} {\bibfield  {journal} {\bibinfo
   {journal} {Scientific reports}\ }\textbf {\bibinfo {volume} {2}} (\bibinfo
  {year} {2012})}\BibitemShut {NoStop}%
\bibitem [{\citenamefont {Strain}\ \emph {et~al.}(2015)\citenamefont {Strain},
  \citenamefont {Lacava}, \citenamefont {Meriggi}, \citenamefont {Cristiani},\
  and\ \citenamefont {Sorel}}]{strain2015}%
  \BibitemOpen
  \bibfield  {author} {\bibinfo {author} {\bibfnamefont {M.~J.}\ \bibnamefont
  {Strain}}, \bibinfo {author} {\bibfnamefont {C.}~\bibnamefont {Lacava}},
  \bibinfo {author} {\bibfnamefont {L.}~\bibnamefont {Meriggi}}, \bibinfo
  {author} {\bibfnamefont {I.}~\bibnamefont {Cristiani}}, \ and\ \bibinfo
  {author} {\bibfnamefont {M.}~\bibnamefont {Sorel}},\ }\href {\doibase
  10.1364/OL.40.001274} {\bibfield  {journal} {\bibinfo  {journal} {Optics
  Letters}\ }\textbf {\bibinfo {volume} {40}},\ \bibinfo {pages} {1274}
  (\bibinfo {year} {2015})}\BibitemShut {NoStop}%
\end{thebibliography}%
\end{document}